\documentclass{aa}
\usepackage{graphicx,txfonts}
\usepackage{natbib}
\usepackage{color}
\usepackage{bm}
\usepackage{hyperref}
\usepackage{braket}
\usepackage{mathrsfs}
\usepackage{multirow}
\usepackage{ulem}

\usepackage[font=small,labelfont=bf,justification=justified]{subcaption}
\definecolor{mygreen}{rgb}{0,0.6,0}
\definecolor{mygray}{rgb}{0.95,0.95,0.95}
\definecolor{mymauve}{rgb}{0.58,0,0.82}

\begin{document}

\title{Layers of electron captures in the crust of accreting neutron stars}
\author{L. Suleiman\inst{1,2,3} \and J.L. Zdunik \inst{2} \and P. Haensel \inst{2}}

\institute{Nicholas and Lee Begovich Center for Gravitational Wave Physics and Astronomy, California State University Fullerton, Fullerton, California 92831, USA. 
 \and Nicolaus Copernicus Astronomical Center, Polish Academy of Sciences, 
Bartycka 18, PL-00-716 Warszawa, Poland
\and Laboratoire Univers et Th\'{e}ories, Observatoire de Paris, Universit\'{e} PSL, Universit\'{e} Paris Cit\'{e}, CNRS, F-92190 Meudon, France}

\date{Received xxx Accepted xxx}

\abstract{
{The accumulation of accreted matter onto the neutron star surface triggers exothermic reactions in the crust. The heat released as a result influences the luminosity exhibited by the X-ray transient. The most common approach to the kinetics of exothermic reactions in the crust of accreting neutron stars is to consider an infinite reaction rate. Here,}
{we investigate accretion-related heat release in the accreted outer crust of a neutron star by including a time-dependent accretion cycle and experimentally based reaction rates in the kinetics of electron captures above the reaction threshold}. 
{A simple model was used to compute the zero temperature equation of state of a crust in which two nuclei can coexist. We solved the abundance of parent nuclei as a function of the depth in the star and the time variable using astrophysically motivated features of the accreting system. We calculated the heat release and neutrino loss associated to reactions in the outer crust.}
{
We report the existence of layers in the outer crust, which contain both parent and grand-daughter nuclei of electron captures. The reactions can occur deeper in the shell than the reaction threshold, thus releasing more heat per accreted baryon for a given accretion rate. The electron capture layers continue to exist even when the accretion has stopped.
The}
{heat sources are time- and pressure-dependent in accreting crusts of neutron stars. The total heat released is a function of astrophysical (active and quiescent time) and microscopic (reaction rate) parameters Therefore, we conclude these parameters should be considered individually and carefully for a range of different sources. }
}


\titlerunning{Layers of electron captures}
\authorrunning{L. Suleiman}

\maketitle

%-------------------------------------------------------------------------%
%----------------------------- SECTION 1 ---------------------------------%
%-------------------------------------------------------------------------%
\section{Introduction}

Neutron stars in binary systems are subject to accretion when matter is transferred to them by their companion star. This highly luminous phenomenon is observed especially well in the X-ray band of the electromagnetic spectrum. Accretion can occur through stellar winds or through an accretion disk if the companion star has evolved into a Roche Lobe donor. In the case of accretion by Roche Lobe overflow, accreted material crashes onto the neutron star surface (spinning up the star in the process) when a loss of angular momentum is triggered in the accretion disk. 

The composition and the equation of state of the crust of an accreting neutron star is different from that of the crust of an isolated (catalyzed) neutron star. As material originating from the companion star crashes onto the neutron star envelope, thermonuclear burning occurs \citep{Meisel_2018}. As a result, ashes with a nucleon number of ${A=\;50-110}$ are deposited on the crust surface and progressively pushed deeper in the star. Contrary to a catalyzed crust that is at global equilibrium, for an accreted crust, the element that falls onto the surface is compressed and subject to non-equilibrium and exothermic reactions. The energy released by electron captures and pycnonuclear fusions deposit heat in the crust: this process is called deep crustal heating \citep{Brown1998}. The heat deposited is transported in the star and radiated through the atmosphere. Then, the resulting luminosity exhibited by the star can be observed after the active accretion process has stopped, that is to say, during quiescence. Consequently, models of accretion related heating of the crust can be compared to the thermal relaxation of X-ray transient sources.

The most common approximation to modeling the crust of an accreting neutron star is the fully accreted crust approximation, where the original crust (i.e., the crust present before any accretion has taken place) is neglected. In this approximation, the crust is solely made of accreted material and the heat sources are always located the same in the crust. Partially accreted crusts are considered when the amount of accreted matter is small enough that the original crust must be studied while it is being compressed under accreted material (more details on this topic are given in \cite{Suleiman2022} and references therein). The usual approach to the crust equation of state and composition is the single nucleus model \citep{Haensel2008, Fantina2018,Potekhin2021}. The role of neutron diffusion was investigated recently \citep{Shchechilin2022}, but it was found that it does not significantly affect the heat release of the outer crust \citep{Potekhin2023}.

Electron captures on an even-even parent nucleus $(A,Z)$, with $A$ as the nucleon number and $Z$ as the proton number,  occur according to the following pair of reactions: 
\begin{align}
    &(A, Z) + e^- \to (A, Z - 1) + \nu \hspace{0.3cm} \text{slow},\; \label{eq:ecSlow}\\ 
    &(A, Z - 1) + e^- \to (A, Z -2) + \nu \hspace{0.3cm} \text{fast} \label{eq:ecfast},\;
\end{align}
with $\nu$ denoting the electronic neutrinos, $(A, Z -1)$ the daughter nucleus, and $(A, Z -2)$ the grand-daughter nucleus. The first electron capture is slow, whereas the second electron capture can be considered instantaneous due to the nuclear energy drop  when passing from an odd-odd nucleus $(A,Z-1)$ to an even-even $(A,Z-2)$ one. Therefore, the kinetics of the pair of reactions is dominated by the first electron capture. The prevailing approach to the kinetics of electron captures in deep crustal heating  is that all nuclei in an infinitesimally small piece of accreted matter are instantaneously changed from parent to daughter nuclei once the pressure threshold of the reaction has been reached \cite{Sato1979, Haensel1990, Haensel2003,Haensel2008, Fantina2018, Shchechilin2019, Chugunov2020, Shchechilin2022}; in the following, this approximation is referred to as the "instantaneous approach."

Full reaction networks beyond the one-component approach were discussed in \cite{Gupta2007, Gupta2008, Steiner2012, Lau2018, Schatz2022}. \cite{Shchechilin2021} found that complete reaction networks lead to a total heat deposited similar to the heat deposited in the "instantaneous approach". The impact of time-dependent accretion on the heat release subsequent to the pycnonuclear fusion of $^{34}$Ne in the inner crust while considering a finite reaction rate and transient accretion model was studied in \cite{Yakovlev2006}. However, the physical processes leading to the reactions are completely different from those considered in this paper, as we study electron captures. 
Finite reaction rates of electron captures in the crust of accreting neutron stars were discussed in \cite{Bildsten1998, Ushomirsky2000}: their study which includes temperature effects, focused solely on stationary accretion and does not include nucleus dependence of the half life in electron capture rates. In this paper, we discuss how including the reaction rate of electron captures with a complete nuclei dependence informed by nuclear experiments affects the composition and the heat release in a fully accreted outer crust constructed within the single nucleus model at zero temperature and subject to time-dependent and transient accretion. 

In Section~\ref{sec:method}, we present the simple approximation taken to model the behavior of a piece of matter with a mixture of parent and grand-daughter nuclei in the outer crust of a neutron star in an accreting binary system. The equation of state followed by the mixture of parent and grand-daughter nuclei is derived in Sect.~\ref{sec:eos}. Transitions between the different shells of the outer crust in this approach are established in Sect.~\ref{sec:Mehtod/compo}. Section~\ref{sec:Mehtod/rate} details the derivation of electron capture reaction rates and the equation to solve the parent nuclei abundance is presented in Sect.~\ref{sec:method/continuity}. Astrophysically motivated cycles of accretion alternating active and quiescent phases, as well as values of accretion rates are discussed in Sect.~\ref{sec:astro}. The heat release and neutrino loss related to the parent nuclei abundance is derived in Sect.~\ref{sec:heat} and Sect.~\ref{sec:neutrino}. Results for the parent nucleus abundance are presented in Sect.~\ref{sec:results} for the shells of an $^{56}$Fe ashes accreted outer crust for various scenarios of the accretion cycle. In addition, the results for the heat released as well as the neutrino loss are presented in this section and compared to the results in the instantaneous approach.

%-------------------------------------------------------------------------%
%----------------------------- SECTION 2 ---------------------------------%
%-------------------------------------------------------------------------%
\section{Methods}\label{sec:method}

%-------------------------------------------------------------------------%
%-------------------------------------------------------------------------%
\subsection{The accreted outer crust equation of state}\label{sec:eos}

A simple approach is taken to establish the zero-temperature relation between the pressure and the density (equation of state) in the outer crust of an accreted neutron star: a lattice allowing for the mixture of two nuclei is permeated by a gas of electrons. The pressure and energy density of a degenerate gas of electrons are given by
\begin{align}
    P_e &= \frac{(m_ec^2)^4}{(\hbar c)^3} \phi(\xi) \;, \\
    \mathcal{E}_e &=\frac{(m_ec^2)^4}{(\hbar c)^3} \chi\left( \xi \right) \;, 
\end{align}
with $\hbar$ as the reduced Planck constant and $c$ as the light velocity such that ${\hbar c = 197.33}$\;MeV\;fm. The electron Fermi momentum is $p_{{\rm F}e}=n_e^{1/3}(3\pi^2)^{1/3}\hbar$ and  the relativity parameter is ${\xi=p_{{\rm F}e}/m_e c}$ with $n_e$ and $m_e$, respectively, the number density and mass of the electron (see e.g., \cite{Shapiro1986}). The expression of the electron's pressure and energy density involves the dimensionless functions: 
\begin{align}\label{eq:layercapPhi(x)}
    \phi(\xi) &= \frac{\xi\left( 1+\xi^2\right)^{1/2}  \left(\frac{2}{3}\xi^2 -1 \right) + \ln \left( \xi + \left( 1+\xi^2\right)^{1/2}\right)}{8\pi^2} \;,\\
    \chi(\xi) &= \frac{\xi\left( 1+\xi^2\right)^{1/2}  \left(1+2\xi^2 \right) - \ln \left( \xi + \left( 1+\xi^2\right)^{1/2}\right)}{8\pi^2} \;. 
\end{align}
In the outer crust, below the liquid ocean, electrons are relativistic: $\xi \gg 1$. In the considered reaction layers, the relativity parameter $\xi$ is between 8 (for the first layer), up to  $\sim 40$ (for the deepest one). With an accuracy better than $\sim 1\%$, one can use ultra-relativistic approximation ${\phi(\xi) \simeq \xi^4/ (12 \pi^2)}$.

An ion  lattice correction to the pressure, denoted $P_{\rm lat}$, is added to the rigid electron background pressure by considering  nuclei of number density $n_{I}$. The volume per nucleus is $1/n_I$ and the volume of a sphere centered at a nucleus (the ion sphere) has the radius  ${a_I=(4\pi/3 \; n_I)^{-1/3}}$. The  
number density of nuclei $n_I$ is related to the electron density, $n_e$, by
\begin{equation}
    n_e  = n_I \left( X Z_0+ (1-X) Z_2\right) \;
\end{equation}
for the capture of two electrons, where the parent nucleus proton number is denoted $Z_0$ and the grand-daughter nucleus proton number is denoted ${Z_2 = Z_0-2}$.
The quantity $X=N_0/(N_0+N_2)$ designates the nucleus $(A,Z_0)$ abundance, with $N_0$ the number of parent nuclei  $(A,Z_0)$ and $N_2$ the number of grand-daughter nuclei  $(A,Z_2)$. 

The lattice pressure $P_{\rm lat}$ for a mixture of two nuclei is given by 
\begin{equation}
    P_{\rm lat} = -0.3 \bigg(\frac{4\pi}{3} \bigg)^{1/3} n_e^{4/3} \; e^2 \; \mathcal{F}(X)\;, 
\end{equation}
with $e$ the elementary charge and $\mathcal{F}$ a function determined by the linear mixing rule (for details, see Section 2.4.7 of \cite{HPY}), which for a mixture of parent and grand-daughter ions is given by the formula 
\begin{equation}
    \mathcal{F}(X) = \frac{(Z_0-2)^{5/3} + X\big[ Z_0^{5/3} - (Z_0-2)^{5/3} \big]}{Z_0 + 2\;\big( X - 1\big)} \;.
\end{equation}

The relation between the baryon density denoted $n$ and the density of electrons in the linear mixing rule approximation is
\begin{equation}
    % n(P,X) = \frac{n_e(P,X) \; A}{X Z_0 + \big( 1-X \big)\big( Z_0 -2\big)}\;.
    n(P,X) = \frac{n_e(P,X)}{\gamma(X)} \;,
\end{equation}
with
\begin{equation}
    \gamma(X) = \frac{X Z_0 + \big( 1-X \big)\big( Z_0 -2\big)}{A}\;.
\end{equation}
The relation between the pressure and the baryon density in this approximation is given by 
\begin{equation}
     P = \frac{(m_ec^2)^4}{(\hbar c)^3} \phi\left( \alpha(X) n^{1/3} \right) + \beta(X)n^{4/3} \;, 
\end{equation}
with
\begin{align}
    \alpha(X) &= \frac{ (3\pi^2)^{1/3} \hbar c}{m_e c^2 } \big(\gamma(X)\big)^{1/3} \;, \label{eq:alphaLMR}\\
    \beta(X) &= -0.3 \bigg(\frac{4\pi}{3}\bigg)^{1/3} e^2 \mathcal{F}(X) \big(\gamma(X)\big)^{4/3}\;.
\end{align}
The chemical potential of a degenerate gas of electrons in a lattice is given by
\begin{equation}
    \mu_e =  \sqrt{n^{2/3} [\alpha(X) m_ec^2]^2 + (m_ec^2)^2} + 4 \frac{\beta(X)}{\gamma(X)} n^{1/3} \;.
\end{equation}
The energy of the nucleus $(A,Z)$ in the heavy nucleus approximation is given by its nuclear mass 
\begin{equation}
    \mathcal{M}(A,Z)c^2 = \mathcal{M}_{\rm at} (A,Z) c^2- Zm_e c^2 + B_{\rm el}(Z) + E_{\rm exc}(A,Z),\;
\end{equation}
with $\mathcal{M}_{\rm at}(A,Z)$ as the atomic mass of the ground state nucleus $(A,Z)$ and $B_{\rm el}(Z)$ the binding energy of $Z$ electrons, as reported in \cite{Lunney2003}. 
The total energy density is given by the formula
\begin{equation}
    \mathcal{E}=X \frac{\mathcal{M}(A,Z)c^2}{A}n+(1-X)\frac{\mathcal{M}(A,Z-2)c^2}{A}n+\mathcal{E}_{\rm lat}+\mathcal{E}_e \;,
\end{equation}
where the lattice energy is $\mathcal{E}_{\rm lat}=3P_{\rm lat}$. 
% and 
% \begin{equation}
%     \mathcal{E}_e=\frac{(m_ec^2)^4}{(\hbar c)^3} \chi\left( \alpha(X) n^{1/3} \right) \;.
% \end{equation}

%-------------------------------------------------------------------------%
%-------------------------------------------------------------------------%
\subsection{Composition of the outer crust}\label{sec:Mehtod/compo}

%-------------------------------------------------------------------------%
\begin{table*}
    \caption{Pressure, $P_{\rm th}$, density, $n_{\rm th}$, and energy threshold, ${W_1, W_2}$, of the four pairs of electron captures in the outer crust made of $^{56}$Fe ashes, calculated in the framework of the linear mixing rule approximation for the equation of state discussed in Section~\ref{sec:eos}.}
    \centering
    \begin{tabular}{|c|cccc|}
        \hline 
        Reaction & $P_{\rm th}$ (MeV/fm$^3$) & $n_{\rm th}$ (fm$^{-3}$) & $W_1$ (MeV) & $W_2$ (MeV) \\
        \hline \hline
        $Z = 26 \to 24$ & $4.13 \times 10^{-7}$ & $8.42 \times 10^{-7}$ & 4.31 & 2.13 \\ \hline
        $Z = 24 \to 22$ & $1.04 \times 10^{-5}$ & $1.02 \times 10^{-5}$ & 9.61 & 7.27 \\ \hline
        $Z = 22 \to 20$ & $5.28 \times 10^{-5}$ & $3.73 \times 10^{-5}$ & 14.42 & 12.51 \\ \hline
        $Z = 20 \to 18$ & $2.40 \times 10^{-4}$ & $1.28 \times 10^{-4}$ & 21.09 & 19.73 \\ \hline
    \end{tabular}
    
    \label{tab:propec}
\end{table*}
%-------------------------------------------------------------------------%

In this paper, the composition of the outer crust is established starting from the fully accreted crust approximation. The crust of the neutron star is entirely composed of accreted matter. The boundaries of shells are defined by the pressure threshold of reactions. We consider the five shells of an accreted outer crust made originally of $^{56}$Fe ashes from thermonuclear flashes. Nuclei  in the entire outer crust have the nucleon number ${A=56}$ because pycnonuclear fusions are not allowed  there. The proton numbers from the neutron star crust surface to the outer and inner crust transition are ${Z=26,\; 24,\; 22,\; 20, \; 18 }$, namely\;: there are four pairs of electron captures that occur during the accretion process onto a fully accreted  outer crust. 

The composition of the outer crust is determined either by the table of experimentally measured nuclei mass table AME2020 presented in \cite{ame2020}, or by the theoretical model HFB-21 presented in \cite{Goriely2010} if masses of nuclei are not available from experimental data (it is the case for $Z=19,18$). The nuclear masses that we use here, as well as their $\beta$-decay schemes, are  in vacuum. This stems from the fact that the electron gas does not influence the structure of the nuclei involved, including their excited states.

In the reactions studied in this paper, we only consider the excited state of the daughter nuclei of the first electron capture, with an excitation energy of $E_{\rm exc}(A,Z-1)$. Actually, it is only the first electron capture in the first shell in the crust that involves an excited state of the nucleus  $^{\rm 56}{\rm Mn}$, with an excited energy of 110\;keV, see Appendix~\ref{sec:appendixTec}. The pressure threshold and consequently the boundaries of the five shells of the outer crust are defined by the energy threshold, $W_1$ and $W_2$, for the first and second electron capture, respectively, and is given by:
\begin{align}
    W_1 &= \mathcal{M}_{\rm at}(A,Z-1) c^2 - \mathcal{M}_{\rm at}(A,Z)c^2 + m_ec^2 \\
    & \hspace{0.5cm} + B_{\rm el}(Z-1) - B_{\rm el}(Z) + E_{\rm exc}(A,Z-1)\;, \nonumber \\
    W_2 &= \mathcal{M}_{\rm at}(A,Z-2) c^2 - \mathcal{M}_{\rm at}(A,Z-1)c^2 + m_ec^2 \\
    & \hspace{0.5cm} + B_{\rm el}(Z-2) - B_{\rm el}(Z-1) \;. \nonumber 
\end{align}
The atomic mass $\mathcal{M}_{\rm at}(A,Z)$ is extracted from experimental data for nuclei ${Z=[26-20]}$ and calculated, using a theoretical approach for $Z=19,18$. The pressure and density threshold denoted $P_{\rm th}$ and $n_{\rm th}$, respectively, defining the transition between each shell is determined by solving ${\mu_e(X=1,n_{\rm th}) = W}$. In Table~\ref{tab:propec}, we present the transition pressures and densities, as well as the energy threshold of the first and second electron captures, respectively, denoted $W_1$ and $W_2$. Our approach is different than the single-nucleus model considered in \cite{Fantina2018} in two respects. On the one hand, experimentally measured nuclei masses used in \cite{Fantina2018} are taken from the AME2016 \cite{ame2016} data table and we used the updated AME2020 data table in our calculations; this is the main source of discrepancy (for the first transition pressure $P_{\rm th}(Z=26)$, including the excitation energy plays a significant role). On the other hand, we do not require a simultaneous transition of all nuclei $(A,Z)\to (A,Z-1)$ and, as a result, the threshold pressure in our model is slightly different (for different treatment of threshold pressure for two models, see \cite{Chamel2015,Chamel2016}).

%-------------------------------------------------------------------------%
%-------------------------------------------------------------------------%

\subsection{Rates of $\beta$-decays  and electron captures} \label{sec:Mehtod/rate}

The reaction rate of $\beta$-decays and electron captures can be established from the Fermi golden rule (originally presented and applied by \cite{Dirac1927}). For details on the derivation of the reaction rates, see Appendix~\ref{AppendixA}. For the $\beta$-decay rate ${(A,Z-1) \to (A,Z)}$ denoted $\mathscr{R}_{\beta}$ the calculation is done in vacuum and exclusively  for allowed type of decays, resulting in the following formula: 
\begin{equation}
\mathscr{R}_{\beta} = \frac{2J_{Z} +1}{2J_{Z-1} +1 } |\mathscr{M}_{\beta}|^2 \frac{4 m_e^5c^4}{(2\pi)^3 \hbar^7} \; f \;, 
\label{eq:betaRate}
\end{equation}
with $J_Z$ as the spin of nucleus $(A,Z)$ and 
where the factor $f$ is the available momentum space factor; notably, the matrix element, $\mathscr{M}_{\beta}$, does not depend on the nuclei spins.

The electron capture rate on   nuclei immersed in the Fermi sea of electrons ${(A,Z) \to (A,Z-1)}$ is denoted by $\mathscr{R}_{\rm ec}$  and can then be expressed as:
\begin{equation}
    \mathscr{R}_{\rm ec} = \frac{2J_{Z-1} +1 }{2 (2J_{Z} +1)} |\mathscr{M}_{\rm ec}|^2 \frac{4 m_e^5c^4}{(2\pi)^3 \hbar^7}\mathscr{G}, \; \label{eq:ecRate}
\end{equation} 
where $\mathscr{G}$ is a dimensionless available momentum space factor, given by 
\begin{align}
     \mathscr{G}(\bar{E}_F, \bar{W}) &= {\int_{\bar{W}}^{\bar{E}_F} \bar{E} \sqrt{\bar{E}^2 -1}\; (\bar{E} - \bar{W})^2  {\rm d}\bar{E}}, \nonumber \\ 
     &= \Big[ \mathscr{G}_{b}(\bar{E}_F, \bar{W}) - \mathscr{G}_{ b}(\bar{W}, \bar{W})\Big] \;, 
\end{align}
with ${\bar{E}_F= E_F /(m_ec^2)}$, ${\bar{W}= W /(m_ec^2)}$ and $E_F$ the Fermi energy of electrons. For the $\beta$-decay reactions, the factor $f$ is ${\propto \mathscr{G}_b(\bar{W}_1,\bar{W}_1) - \mathscr{G}_b(1,\bar{W}_1)}$\; and often includes corrections for the Coulomb interaction. The generalized function, $\mathscr{G}_{b}$, can be obtained via an analytical formula, first presented in \cite{Frank1962}\footnote{Note, however, that the integrand in Eq.(4) in \cite{Frank1962} has to be corrected by replacing $\varepsilon^2$ in the brackets by $\varepsilon$.} as: 
\begin{align}
    \mathscr{G}_{b}(x,w) &= \frac{\sqrt{x^2-1}}{60}  \bigg(   x^3(12x-30w) -4x^2(1-5w^2) \label{eq:Fbfunction}  \\
    &+15w x-20w^2-8 \bigg) +\frac{1}{4}\ w \log \left( \sqrt{x^2-1}+x\right) \nonumber \;.
\end{align}

In the ultra-relativistic approach ($\bar E_F\ge \bar W \gg 1$), the expansion up to the first order in $\bar W^{-2}$ yields the approximation $ \mathscr{G}\approx \mathscr{G}_{\rm ultra}$: 
\begin{align}\label{eq:Gultra}
     \mathscr{G}_{\rm ultra}(E_F, W) &= \frac{\bar{W}^5}{3} \bigg( 
     \frac{E_F}{W} -1\bigg)^3 \bigg[ 1 - \frac{1}{2\bar{W}^2} + \frac{3}{2} \bigg(\frac{E_F}{W} -1 \bigg) \nonumber \\ 
     & \hspace{1.5cm}+ \frac{3}{5} \bigg(\frac{E_F}{W} -1\bigg)^2\bigg] \;.
\end{align}

For an electron Fermi energy close to the reaction threshold energy (${E_F-W\ll W}$), the leading order expression is:
\begin{equation} \label{eq:ultraAndcloseTh}
  \mathscr{G}_{\rm ultra}(E_F, W)=  \frac{\bar{W}^5}{3} \bigg( 
     \frac{E_F}{W} -1\bigg)^3 \;, 
\end{equation}
here, we also refer to  Eq.~(4) and the zero temperature limit of Eq.~(6) in \cite{Bildsten1998}. We note that the ultra-relativistic approximation is used in this paper only to make analytical estimates, the full numerical solution makes use of Eq.~\eqref{eq:Fbfunction} to formulate the reaction rate.

Using charge conjugation symmetry as well as time reversal symmetry, we can show that the matrix element for the electron capture and its corresponding $\beta$-decay in the above mentioned approximations are equal. Therefore, we can make use of the experimental data available for $\beta$-decay reactions to compute the reaction rate of the first electron capture per parent nucleus:
\begin{align}
     \mathscr{R}_{\rm ec} &= \frac{\ln(2)}{f t_{1/2}} \frac{2J_{\rm ec; Z-1}+1}{2 (2J_{\rm ec; Z}+1)} \frac{2J_{\beta; Z-1}+1}{2J_{\beta; Z}+1}\mathscr{G}(\bar{E}_F, \bar{W}_1) ,\\
     &\equiv \frac{1}{\tau_{\rm ec}} \mathscr{G}(\bar{E}_F, \bar{W}_1) \;,
\end{align}
where $J_{\rm ec;Z}$ and $J_{\beta ;Z}$ are nuclear spins  for the electron capture and its corresponding $\beta$-decay, respectively. The quantity $t_{1/2}$ is the half-life of the $\beta$-decay for a specific reaction channel and can be extracted from experimental data or estimated analytically, $\tau_{\rm ec}$ is the electron capture timescale. Electron captures and $\beta$-decays are subject to selection rules \citep{Povh2004} that specify allowed  changes of nuclear spins for nuclei involved in electron captures and $\beta$-decays. We selected the dominant channel in the energy level diagrams presented in \cite{ENSDF} for each reaction. In Appendix~\ref{sec:appendixTec}, we give some details on the calculation of the electron capture rates.  In Table~\ref{tab:timescales}, we present values of $\tau_{\rm ec}$ for the first electron capture involved in the four shells of the outer crust. 

%-------------------------------------------------------------------------%
%-------------------------------------------------------------------------%
\subsection{Continuity equation} \label{sec:method/continuity}

During the active phases of accretion, freshly accreted matter is flowing and sinking towards the core with a velocity: 
\begin{equation}
    v(z) = \frac{\dot{M}}{4\pi R^2 m_B n(z)} \;, 
\end{equation}
with $\dot{M}$ the mass accretion rate, $R$ the total radius of the star, ${m_B}$ the rest mass of a baryon, and $z=R-r$ the proper distance from the surface in the plane-parallel approximation.
It should be noted that even if the accretion stops (quiescent phase), nuclear reactions still occur leading to a slow change of the equation of state in the mixed layer. This results in a shrinking of the reaction layer, and therefore to a certain (low) velocity. In our considerations we use the pressure as an independent variable (in place of $z$ or $r$). For a given piece of matter, the pressure changes due to accretion with the increase proportional to the accreted mass; during quiescent phases, the pressure does not change as the column of matter above is fixed when the reactions take place.

The most important quantity introduced in the previous section is the parent nucleus abundance, $X$. The layers of electron captures are defined as the thickness in pressure, for which $X\in [0-1]$. The product of the parent nucleus abundance, $X$, with the baryon density corresponds to the number density of parent nuclei, and is governed by the continuity equation. This equation is written in the local frame with $\tau$ and $z$ corresponding to the proper time and proper distance in the neutron star crust. In what follows, we use the Newtonian approach
\begin{equation}
    \frac{\partial P}{\partial r} = -\frac{GM}{R^2} \rho \;, \label{eq:Poisson}
\end{equation}
where $\rho=\epsilon/c^2 \approx n m_B$, and $M$ is the total mass of the star; in the following the mass and the radius of the star are chosen to be $1.4$\;M$_{\odot}$ (solar mass) and 11\;km, respectively. The continuity equation is therefore given by: 
\begin{align}\label{eq:ce}
    \frac{\partial}{\partial t} \ln \big( n\big( X(P,t),P \big) \; X(P,t) \big) &+  \frac{1}{\tau_{\rm acc}(t)}\frac{\partial \ln X(P,t) }{\partial \tilde{P}} \nonumber \\
    &= -\mathscr{R}_{\rm ec} \big(X(P,t), P\big)\;,
\end{align}
with $\tilde{P} = P/P_{\rm th}$, with $n=n(P,X)$ and $X=X(P,t)$, and $\tau_{\rm acc}$ is the accretion timescale,
\begin{equation}
    \tau_{\rm acc}(t) = \frac{4\pi R^4 P_{\rm th} }{GM\dot{M}(t)} \;.
    \label{eq:taua}
\end{equation}
Typical values of the accretion timescale are presented in Table~\ref{tab:timescales}. In the general relativity approach, with a metric ${ds^2=e^{2\phi}dt^2-e^{2\lambda}dr^2-d\Omega^2}$, the continuity equation has the same form as in the Newtonian one in the local frame with coordinates ($te^{\phi}$, $re^{\lambda}$), and the boundary condition at the surface is ${e^{\phi(R)+\lambda(R)}=1}$. By choosing the pressure, $P$, as the independent variable instead of $z$, the relativistic equivalent of Eq.~\eqref{eq:ce} would include $\dot{M}$ multiplied by a factor $e^{\phi(R)}=(1-2GM/Rc^2)^{1/2}$ in the expression of the accretion timescale (Eq. \eqref{eq:taua}); in our case ($M=1.4$\;M$_\odot$, $R=11$\;km), this factor is $\sim$\;0.8. The mass accretion rate would be defined as $\dot M=\dot N/m_B$, with $\dot N=dN/dt$, as measured by a distant observer. In the rest of the paper, we neglect relativistic effects and treat the Newtonian approach of the continuity equation.

%-------------------------------------------------------------------------%
\begin{table}
    \caption{Electron capture timescale and accretion timescale for the first electron capture for the four pairs in the outer crust made of $^{56}$Fe ashes. The accretion timescale is calculated for a 1.4\;M$_{\odot}$ neutron star mass and 11\;km radius considering a constant accretion rate of ${10^{-8}}$\;M$_{\odot}$ per year.} 
    \centering
    \begin{tabular}{|c|cc|}
        \hline 
        Reaction & $\tau_{\rm ec}$ (in s) & $\tau_{\rm acc}$ (in s) \\
        \hline \hline
        $Z = 26 \to 25$ & $8.67 \times 10^{6}$ & $9.50 \times 10^{7}$\\ \hline
        $Z = 24 \to 23$ & $1.37 \times 10^{4}$ & $2.40 \times 10^{9}$ \\ \hline
        $Z = 22 \to 21$ & $7.34 \times 10^{3}$ & $1.21 \times 10^{10}$\\ \hline
        $Z = 20 \to 19$ & $1.59 \times 10^{5}$ & $5.67 \times 10^{10}$ \\ \hline
    \end{tabular}
    \label{tab:timescales}
\end{table}
%-------------------------------------------------------------------------%

%-------------------------------------------------------------------------%
%-------------------------------------------------------------------------%
\subsection{Astrophysically motivated accretion cycle}\label{sec:astro}

We intend to study a realistic cycle of accretion (sequence of active and quiescent phases) motivated by X-ray observations of sources in accreting low-mass X-ray binaries. Several sources have been observed alternating between active accretion stages lasting from weeks to years and quiescent stages lasting from months to decades. The information on characteristic duration of active accretion can be extracted from the observations of X-ray outbursts (a lasting surge in luminosity) of sources EXO $0748$–$676$ (see \cite{Parikh2020} and references therein), KS $1731$–$260$ (see \cite{Merritt2017} and references therein), XTE J$1701$–$462$ \citep{Fridriksson2010}, and IGR J$17480$-$2446$ (see \cite{Ootes2019} and references therein). This lasted  24 years, 12.5 years, 1.6 years, and 10 weeks, respectively; all four sources are now in quiescence. The source MXB $1659$-$29$ (see \cite{Parikh2019} and references therein) presented two well monitored outbursts lasting, respectively, 2.5 and 1.7 years, interspersed by a quiescent period of 14 years. This source was first observed in 1976 during an outburst estimated to last between 2 and 2.5 years as well. Unless otherwise stipulated, in this paper we study the following cycle: an active phase period, $t_a$, of four years and a quiescent period of ${t_q = 10 t_a}$.  

The average accretion rate during active phases were estimated for the sources XTE J$1701$-$462$ \citep{Fridriksson2010} and IGR J$17480$-$2446$ \citep{Degenaar2011b} to be ${1.7 \times 10^{-8}}$~\;M$_{\odot}$ per year and ${3 \times 10^{-9}}$~\;M$_{\odot}$ per year, respectively. The observation of X-ray bursts indicate a luminosity variability in the active phase from which the accretion rate as a function of time can be inferred, however, we do not intend to discuss this variability;  rather, we aim to use a mean value of the accretion rate in the active phase. The active accretion rate, denoted $\dot{M}_a$, is described by an exponential onset and offset such that: 
\begin{equation}\label{eq:mdotForm}
    \dot{M}_a(t) = \dot{M}_{\rm max} \bigg( \frac{1}{1+e^{a (t-t_{\rm a} + t_{\rm o})}} + \frac{1}{1+e^{-a (t- t_{\rm o})}} - 1 \bigg) \;, 
\end{equation}
with $\dot{M}_{\rm max}$ the accretion rate outside the onset and offset of active accretion and $a$ a constant controlling the steepness of the $\dot M$ increase and decrease during onset and offset. For $a\simeq 118$ (in yr$^{-1}$), the onset and offset are very steep. We consider the accretion is switched on as soon as $\dot{M}$ reaches $10^{-10}\dot{M}_{\rm max}$. At $t_o$, half of the maximum value of the accretion rate  $\dot{M}_{\rm max}/2$ is reached. After $t_o$, because $a$ is so large and the onset so steep, we quickly reach $\dot{M}_{\rm max}$ to a very high precision. Unless otherwise stipulated, in this paper we set ${\dot{M}_{\rm max}=10^{-8}}$\;M$_{\odot}$ per year, a reasonable value with regard to observed mean accretion rate during active accretion, and ${t_o = 0.05t_a}$.

The number of accreted baryons during an accretion cycle denoted $N_b$ is defined by the time integral of the accretion rate during the cycle
\begin{equation}
    N_b = \int_0^{t_a+t_q} \frac{\dot{M}_a(t)}{m_B} \; {\rm d}t \;. \label{eq:accBaryons}
\end{equation}
Given that we consider a finite reaction rate, the number of accreted nuclei over one accretion cycle is not \textit{a priori} the same as the number of nuclei undergoing electron captures. Therefore, we define $\dot{N}_r$ the rate of baryons involved in an electron capture per unit time and per unit volume
\begin{equation}
    \dot{N}_r(P, t)  = \mathscr{R}_{\rm ec}(P, X) X(P,t) \;.\label{eq:reactiveBaryonsDot}
\end{equation}
The number of reactive baryons per unit time in the shell gives
\begin{equation}
    N_r(t) = \frac{4 \pi R^4 P_{\rm th}}{GMm_B} \int_{1}^{\tilde{P}_0} \mathscr{R}_{\rm ec}(\tilde{P}, X) X(\tilde{P},t){\rm d}\tilde{P} \;, 
\end{equation}
where ${\tilde{P}_0 = P(X=0)/P_{\rm th}}$\;. The number of reactive baryons in the shell over the cycle of accretion
\begin{equation}
    N_r = \int_0^{t_a+t_q}N_r(t) {\rm d}t\;. \label{eq:reactiveBaryons}
\end{equation}

%-------------------------------------------------------------------------%
%-------------------------------------------------------------------------%
\subsection{Heat release from electron captures}\label{sec:heat}

We consider the first electron capture presented in Eq.~\eqref{eq:ecSlow}, while neglecting the neutrino heat loss (cooling). We also consider that processes under consideration proceed at constant pressure and temperature in a piece of matter. Following the derivation of Chapter X in \cite{Landau1969}, we define an elementary reaction ${e^- + (A,Z) \longrightarrow (A,Z-1)}$, with the coefficients ${\nu_e=\nu_{A,Z}=1}$ and ${\nu_{A,Z-1}=-1}$. The number of electron captures in the time interval ${\rm d}t$ is denoted ${\delta N_{\rm ec}}$; the changes of numbers of particles $i$ are ${\delta N_i=-\nu_i\delta N_{\rm ec}}$, while the thermodynamic potential $\Phi$  changes by ${\delta \Phi=\sum_i\delta N_i\mu_i=-\sum_i\nu_i\mu_i \delta N_{\rm ec}}$. The heat released at constant pressure and temperature is (Chapter X in \cite{Landau1969}): 
\begin{equation}
    \delta Q_P=T\delta S=-T^2\left(\frac{\partial}{\partial T}\frac{\delta \Phi}{T}\right)\;, 
\end{equation}
with $T$ the temperature and $S$ entropy. In our approximation, $\mu_i$ values do not depend on $T$ (electrons are fully degenerate and nuclear masses do not depend on $T$), so that
\begin{equation}
    {\delta Q_P}=-\sum_i\nu_i\mu_i \delta N_{\rm ec} \;.  
\end{equation}
The heating rate is therefore
\begin{equation}
    {\dot{Q}_P}= T\dot{S}=-\sum_i\nu_i\mu_i {\dot{N}_{\rm ec}}~.   
\end{equation}

The heat release per nucleus for the first and second electron capture of the pair, denoted respectively $q_1$ and $q_2$, neglecting neutrino heat loss, are 
\begin{align}
    q_1(P,X) &= \mathcal{M}(A,Z)c^2 - \mathcal{M}(A,Z-1)c^2 + \mu_e(P,X) \;, \\
    q_2(P,X) &= \mathcal{M}(A,Z-1)c^2 - \mathcal{M}(A,Z-2)c^2 + \mu_e(P,X) \;.
\end{align}
The heat released by the pair of electron captures per nucleus is denoted $q$ and is given by
\begin{align}\label{eq:QabsMixed}
    q(P,X) = 2 \mu_e(P,X) - ( W_1 + W_2) + E_{\rm exc} \;, 
\end{align}
with $E_{\rm exc}$ the excitation energy of the first electron capture daughter nucleus; note that except for $^{56}$Mn$^*$, we only consider ground-state to ground-state reactions. Therefore, the heating rate per unit volume is given by 
\begin{equation}\label{eq:layercapHeatingRate}
    \dot{q}(P,t) = q(P,X) \mathscr{R}_{\rm ec}(P,X) \frac{n(P,X)}{A} X(P,t)\;.
\end{equation}

The heat released in the layer of electron captures per unit time is
\begin{align}
    \dot{\mathcal{Q}}(t) &= \frac{4 \pi R^4 P_{\rm th}}{GMm_B} \int_{1}^{\tilde{P}_0} 
    \frac{ \dot{q}(\tilde{P},t)}{n(\tilde{P},X)}{{\rm d}\tilde{P}} \;.
\end{align}

The total heat released in a layer per one accreted nucleon during one accretion cycle is
\begin{equation}\label{eq:totalQ}
    Q = \frac{1}{N_b}\int_0^{t_a+t_q}\dot{\mathcal{Q}}(t) {\rm d}t \;, 
\end{equation}
with $N_b$ the number of accreted nucleons during this cycle.
%-------------------------------------------------------------------------%
%-------------------------------------------------------------------------%
\subsection{Neutrino energy loss}\label{sec:neutrino}
In the assessment of the heat release in the outer crust of a neutron star in an accreting binary system, the neutrinos must be taken into consideration: in conditions relevant for adult neutron stars, neutrinos escape the system taking away energy.

Similarly to Eq.~\eqref{eq:QabsMixed}, we can define the energy allocated to neutrinos in the pair of electron captures per nucleus
\begin{align}\label{eq:qnu}
    q_{\nu}(P,X) &= m_ec^2 \left( \frac{\mathscr{G}_{\nu} (\bar{E}_F,\bar{W}_1)}{\mathscr{G}(\bar{E}_F,\bar{W}_1)} +  \frac{\mathscr{G}_{\nu} (\bar{E}_F,\bar{W}_2)}{\mathscr{G}(\bar{E}_F,\bar{W}_2)} \right) \;.
\end{align}
The energy of neutrinos per electron capture and per parent nucleus, in the units of $m_e c^2$,  denoted $\mathscr{G_{\nu}}$, is given by
\begin{align}
     \mathscr{G}_{\nu}(\bar{E}_F, \bar{W}) &= \int_{\bar{W}}^{\bar{E}_F} \bar{E} \sqrt{\bar{E}^2 -1}\; (\bar{E} - \bar{W} )^3  {\rm d}\bar{E} \nonumber \\ 
     & = \Big[ \mathscr{G}_{\nu b}(\bar{E}_F, \bar{W}) - \mathscr{G}_{\nu b}(\bar{W}, \bar{W})\Big] \;;
\end{align}
the function $\mathscr{G}_{\nu b}$ can be obtained analytically as 
\begin{align}
    \mathscr{G}_{\nu b}(x,w) &= \frac{1}{240} \bigg(\sqrt{x^2-1} \Big[ 40x^5-144wx^4 \\
     & +10x^3 (18w^2-1) +16wx^2(3-5w^2) \nonumber\\
     &-15x(1+6w^2)+16w(6+5w^2) \Big] \nonumber\\
     & -15(1+ 6w^2) \Big[\log \left(\sqrt{x^2-1}+x\right) \Big]\bigg) \nonumber \;,
\end{align}
see also \cite{BKogan2001}.

The energy radiated away by neutrinos per unit time and per unit volume $\varepsilon_{\nu}$ is given by 
\begin{equation} \label{eq:enuemission}
    \varepsilon_{\nu}(P,t) =  q_{\nu}(P,X)\mathscr{R}_{\rm ec}(P,X) \frac{n}{A} X(P,t) \;.
\end{equation}
The heat deposited in the layer of electron capture per time unit is defined as the integral over the volume of the difference between the heating rate and the cooling rate due to neutrino emission
\begin{align}
   \dot{\mathcal{H}}(t) &= \frac{4 \pi R^4 P_{\rm th}}{GMm_B} \int_{1}^{\tilde{P}_0} 
   \frac{ \dot{q}(\tilde{P},t) - \varepsilon_{\nu}(\tilde{P},t)}{n(\tilde{P},X)} {\rm d}\tilde P\;.
\end{align}
The total heat deposited per one accreted baryon during one accretion cycle gives 
\begin{equation}\label{eq:totalH}
   H = \frac{1}{N_b}\int_0^{t_a+t_q}\dot{\mathcal{H}}(t) {\rm d}t \;.
\end{equation}

\subsection{Heating from  $(A,Z-1)^*$ decay versus neutrino cooling}\label{sec:excited}
%------------------------------------------------------------
The daughter nucleus $(A,Z-1)$ from the first electron capture in Eq.(1) has odd numbers of neutrons and protons. Therefore, it has a lower energy threshold for the second electron capture, $W_2<W_1$. Moreover, it has numerous excited states above the ground state one. This means that there is an energy range of electrons $W_2<E_e<E_F$, whose capture leads to the excited states $(A,Z-1)^*$. These excited states with energies $0<E_{\rm exc}<E_e-W_2$ de-excite via electromagnetic transition, heating the matter. The multi-component plasma calculations of the accreting crust evolution show that the de-excitation heating balances the neutrino  heat losses calculated in Sect.\ref{sec:neutrino} \citep{Gupta2007}. Therefore, neglecting neutrino losses and using  the heating rate calculated in Sect. \ref{sec:heat} is a reasonable approximation. Such an approximation was used in \citep{Haensel2008,Fantina2018}.

%-------------------------------------------------------------------------%
%----------------------------- SECTION 3 ---------------------------------%
%-------------------------------------------------------------------------%
\section{Results} \label{sec:results}

%-------------------------------------------------------------------------%
%-------------------------------------------------------------------------%
\subsection{Thickness of the layer of electron capture}
%-------------------------------------------------------------------------%
\subsubsection{Stationary solution}

%-------------------------------------------------------------------------%
\begin{figure}
    \centering
    \resizebox{\hsize}{!}{\includegraphics{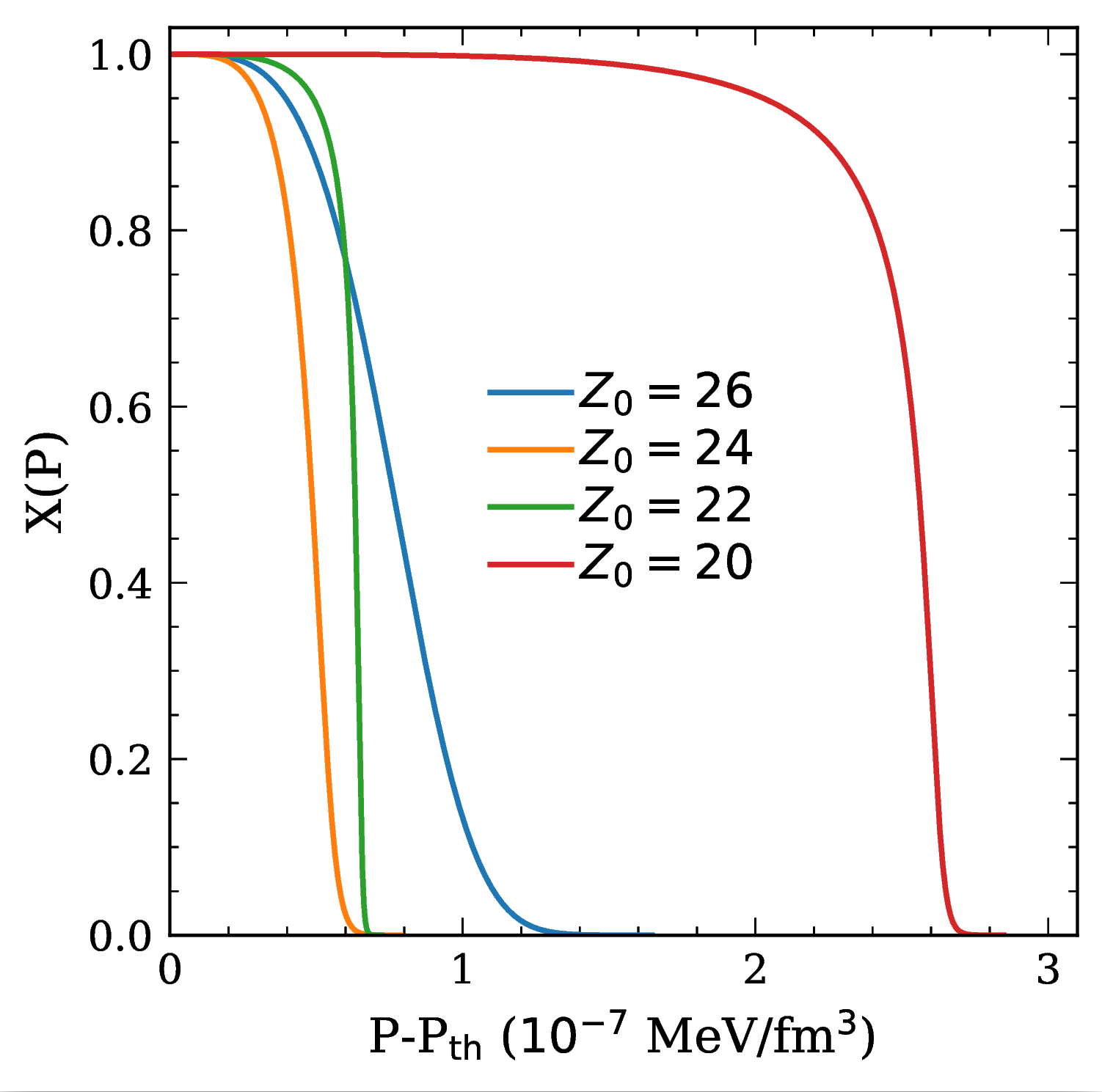}}
    \caption{Stationary solution for the parent nucleus abundance, $X$, as a function of the pressure for the four reactions in the outer crust. }
    \label{fig:stationary}
\end{figure}
%-------------------------------------------------------------------------%

As a first approach to the behavior of the parent nucleus abundance, $X$, in each shell of the outer crust, we establish the solution of the continuity equation for a constant accretion rate; the solution depends only on the pressure. Equation~\eqref{eq:ce} is reduced to an ordinary differential equation treated with a Runge-Kutta numerical approach, with a constant accretion rate of ${\dot{M}= \dot{M}_{\rm max}=10^{-8}}\;{\rm M}_{\odot}$ per year. The boundary condition ${X(P=P_{\rm th}) =1}$ is used: at the pressure threshold, all nuclei are parent nuclei. The accreted material is continuously pushed deeper in the crust by freshly accreted material and the solution does not depend on time. 

When a piece of matter composed of parent nuclei reaches the pressure threshold of the first electron capture, parent nuclei are allowed to transform into daughter (and immediately grand-daughter) nuclei. In the instantaneous approximation, all parent nuclei react at the threshold pressure of the first electron capture. However, when considering finite reaction rates, parent nuclei electron captures take place according to the finite reaction rate and as the piece of matter is still pushed deeper in the crust. Contrary to the instantaneous approximation, in our approach (later referred to as the mixed layer approach), the sinking piece of matter still contains parent nuclei that shall be changed to grand-daughter nuclei at values of the pressure higher than the threshold pressure. The stationary solution corresponds to a particular distribution of the parent nuclei number as a function of the pressure: it appears during active phases of accretion if a steady state has been reached between a continuous supply of parent nuclei (and compression) provided by accretion and the reactions operating on the parent nuclei.

The stationary solution $X_{\rm stat}(P)$ is presented for parent proton number $Z_0 = 26,\;24,\;22,\;20,$ in Fig.~\ref{fig:stationary}. We define the thickness in pressure of the layer of electron capture, that is to say, the region in the shell where both parent and grand-daughter nuclei are found, using the value of the pressure at which the shell is made of only 1\% of parent nuclei. It is denoted ${\Delta P_{\rm layer} = P(X= 10^{-2}) - P_{\rm th}}$\;; this  quantity is also presented in Table~\ref{tab:deltaP} for the stationary solution. For all four pairs of electron captures, the thickness of the layer is on the order of $10^{-7}$\;MeV/fm$^3$. The slope of decrease in the parent abundance for the stationary solution is correlated with the ratio ${\tau_{\rm acc}/\tau_{\rm ec}}$ (see Table~\ref{tab:timescales}): the higher the ratio, the steeper the decrease.

It is important to note that the thickness of the electron capture layer is much smaller than the thickness of the shell composed of the grand-daughter nuclei. The thickness in pressure for all considered shells is $\sim 10^{-5}-10^{-3}$\;MeV/fm$^3$; the reaction layer thickness represents at most one percent of the shell thickness, as said shells are located deeper in the crust and the layer thickness is on the same order of magnitude throughout the outer crust. This implies that no shell of the outer crust with parent number nucleus, $Z'$, is also composed of its shallower neighbor parent nucleus, $Z'-4$.

Using the ultra-relativistic approximation of the electron capture reaction rate presented in Eq.~\eqref{eq:Gultra}, we can provide a simple analytical approximation of the increase in pressure $\delta P$ required to reach the region in which most of the parent nuclei are transformed into grand-daughter nuclei. The stationary form of the continuity equation for ultra-relativistic electrons is written as 
\begin{equation}\label{eq:ultraContinuity}
    \frac{\tau_{\rm ec}}{\tau_{\rm acc}} P_{\rm th} \frac{\partial \ln(X)}{\partial(\delta P)} = -\frac{\bar{W}_1^5}{3} \bigg( \frac{E_F}{W_1} -1\bigg)^3 \;.
\end{equation}
In this approximation, we include only the leading term in Eq.~\eqref{eq:Gultra} - it is valid for $E_F/W_1-1\ll 1$, equivalent to the condition $\delta P/P\ll 1$, (see Eq.~\eqref{eq:ultraAndcloseTh}). This condition is fulfilled for all considered reaction layers but the first one, for which it is on the order of 0.1-0.2 (see Table~\ref{tab:propec},\ref{tab:deltaP}).  Under the assumption that the parent nuclei abundance is close to one, the factor $\alpha$ (see Eq.~\eqref{eq:alphaLMR}) related to the linear mixing rule function is considered to be constant; thus, the right hand side of Eq.~\eqref{eq:ultraContinuity} is independent of $X$. Solving the approximation of the continuity equation in the ultra-relativistic approach (${P\sim E_F^4}$ and ${\delta P/P \simeq 4(E_F/W-1)}$), we obtain the approximate solution to Eq.~\eqref{eq:ultraContinuity} ${X(\delta P)=\exp{[-(\delta P/\delta P_{\rm s})^4],}}$ where $\delta P_{\rm s}$ is e-folding value of $\delta P$ at which $X=1/e$  defined by:
\begin{equation}
    \delta P_{\rm s} =4  P_{\rm th} \bigg(\frac{3}{\bar{W}_1^5}\frac{\tau_{\rm ec}}{\tau_{\rm acc}}\bigg)^{1/4}. \;
    \label{deltaPreac}
\end{equation}
This approximation works well for the a small decrease of $X$: for $X=0.95$, it can be quite accurately determined by 
\begin{equation}
   \delta P(X=0.95) = 0.476\,\delta P_{\rm s}= 2.51  P_{\rm th}  \bigg(\frac{1}{\bar{W}_1^5}\frac{\tau_{\rm ec}}{\tau_{\rm acc}}\bigg)^{1/4},\;
\label{deltaPreac95}
\end{equation}
a quantity presented in Table~\ref{tab:deltaP}.

%-------------------------------------------------------------------------%
\begin{table}
    \caption{Estimation in the ultra-relativistic electron approximation of the thickness, $\delta P (X=0.95) = P(X=0.95) - P_{\rm th}$, and thickness in pressure of the layer, $\Delta P_{\rm layer}$, of electron capture established from the stationary solution.}
    \centering
    \begin{tabular}{|c|cc|}
        \hline 
        Reaction & $\delta P(X=0.95)$\;MeV/fm$^3$& $\Delta P_{\rm layer}$\;MeV/fm$^3$ \\
        \hline \hline
        $Z_0 = 26 $ & $3.86\times 10^{-8}$ & $1.24 \times 10^{-7}$\\ \hline
        $Z_0 = 24$ & $3.18\times 10^{-8}$ & $6.20 \times 10^{-8}$ \\ \hline
        $Z_0 = 22$ & $5.53 \times 10^{-8}$ & $6.71 \times 10^{-8}$\\ \hline
        $Z_0 = 20$ & $2.31 \times 10^{-7}$ & $2.68 \times 10^{-7}$ \\ \hline
    \end{tabular}
    \label{tab:deltaP}
\end{table}
%-------------------------------------------------------------------------%

Similarly to the estimation of the pressure thickness in Eq.~\eqref{deltaPreac}, we can estimate the radius thickness of the electron capture layers without solving the continuity equation as: 
\begin{equation}
    \delta R =4  \frac{R^2 P_{\rm th}}{GM\rho_{\rm th}} \bigg(\frac{3}{\bar{W}_1^5}\frac{\tau_{\rm ec}}{\tau_{\rm acc}}\bigg)^{1/4} \;, 
    \label{deltaRreac}
\end{equation}
with $\rho_{\rm th} = n_{\rm th} m_B $. For a neutron star mass of $M=1.4\;{\rm M}_{\odot}$ and radius $R=11$\;km: the layers for $Z_0=26$, $Z_0=24$, $Z_0= 22,$ and $Z_0=20$ are about 6\;m, $0.4$\;m, $0.2$\;m, and $0.2$\;m thick, respectively. This is much less than the thickness of the shells which in all studied cases are larger than 100\;m (see Eq. (10) in \cite{Suleiman2022}). The maximum size of the reaction layer (corresponding to stationary solution) is significantly smaller than the thickness of the shell also for other, astrophysically relevant accretion rates. The presented numbers correspond to $\dot M = 10^{-8}~{\rm M_\odot/yr,}$ but even a decrease of $\dot M$ by two orders of magnitude only leads to a factor  of $\sim 0.3$ in Eqs.~\eqref{deltaPreac}-\eqref{deltaRreac}. For a small (5\%) decrease in X, we get:
\begin{equation}
    \delta R (X=0.95)= 2.5 \frac{R^2 P_{\rm th}}{GM\rho_{\rm th}} \bigg(\frac{1}{\bar{W}_1^5}\frac{\tau_{\rm ec}}{\tau_{\rm acc}}\bigg)^{1/4}, \; 
    \label{deltaRreac95}
\end{equation}
in the case of which we obtained about 3\;m for the first reaction layer and $\sim 0.1-0.2~\rm{m}$ for deeper reaction layers\footnote{See also similar considerations for the hydrogen electron capture in the NS ocean in \cite{Bildsten1998}, Eq.~(13,14).}.

%-------------------------------------------------------------------------%
\subsubsection{Time-dependent active phase solution for $Z_0=26$}

%-------------------------------------------------------------------------%
\begin{figure*}
    \centering
    \begin{subfigure}[b]{0.45\hsize}
    \resizebox{\hsize}{!}{\includegraphics{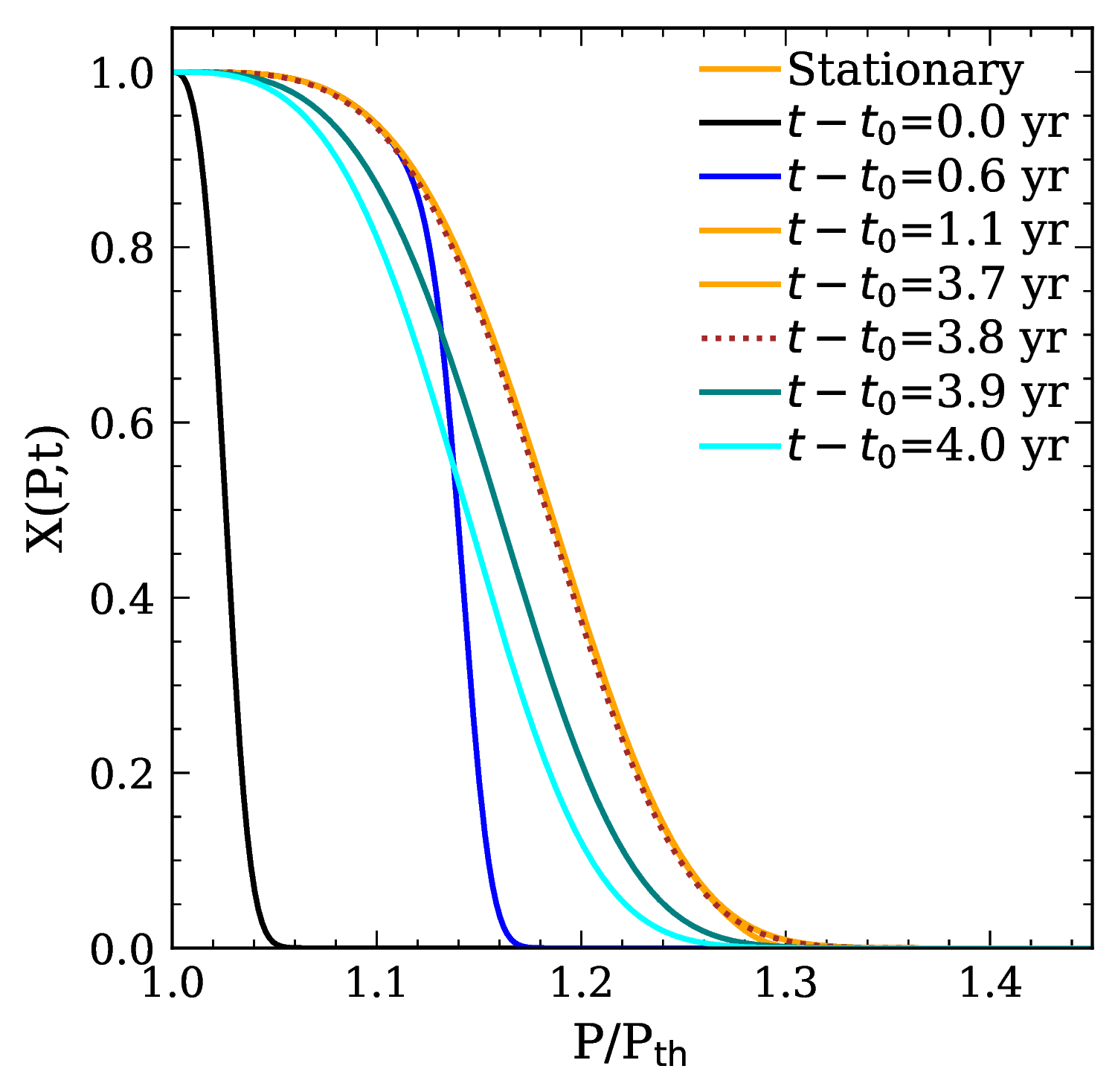}}
    \caption{\large{$\dot{M}_{\rm max} = 10^{-8}$\;M$_{\odot}$ per year.}}
    \label{fig:fullSoluce}
    \end{subfigure}
    \hfill
    \begin{subfigure}[b]{0.45\hsize}
     \resizebox{\hsize}{!}{\includegraphics{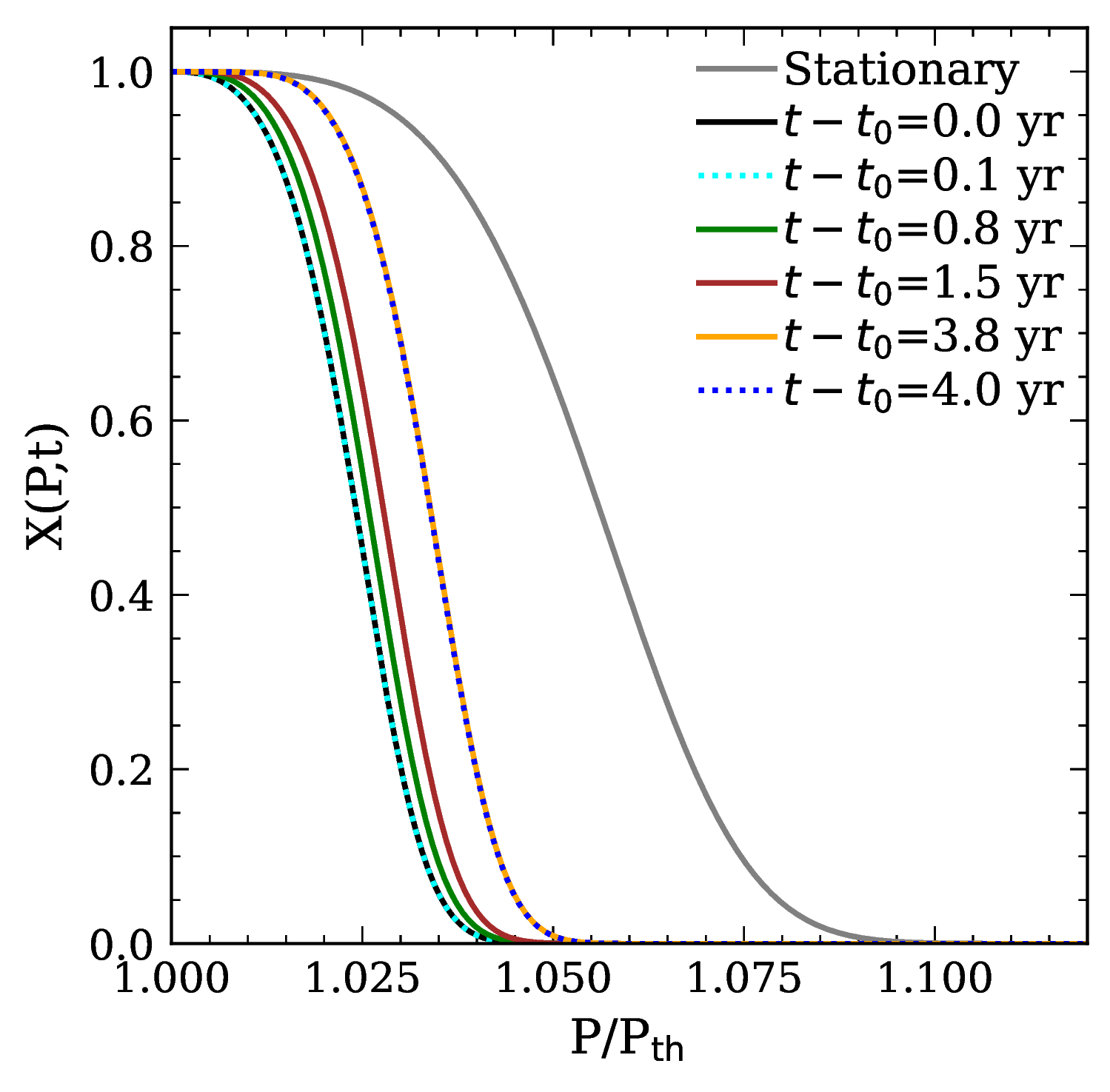}}
    \caption{\large{$\dot{M}_{\rm max} = 10^{-10}$\;M$_{\odot}$ per year.}}
    \label{fig:fullSoluceMdot}
    \end{subfigure}
    \caption{Time-dependent solution, $X(P,t),$ represented as snapshot at various times during the four years of active accretion cycle, starting at $t_0$}. 
\end{figure*}
%-------------------------------------------------------------------------%
To solve the complete continuity equation in Eq.~\eqref{eq:ce} during active accretion, we used a time-dependent accretion timescale with ${\dot{M}_{\rm max} = 10^{-8}\;{\rm M}_{\odot}}$. The following accretion cycle parameters were used in Eq.~\eqref{eq:mdotForm}: $t_a=4$\;years and $t_o=0.2$\; years. The boundary condition is the same as the stationary case and the initial condition is ${X(P,t=0)=X_{\rm stat}(P)}$. This simplistic approach is reasonable if the equation is solved for multiple accretion cycles and we present results for the fourth accretion cycle for the shell with $Z_0=26$. It turned out that after two cycles, we reached a stable solution, in which $X(P,t)=X(P,t+\tau_{cycle}),$ where $\tau_{cycle}=t_a+t_q$ is the full period of our accreting source. After reaching this stable solution all cycles look identical.

Results for the parent nucleus abundance $X(P,t)$ for $Z_0=26$ are presented in Fig.~\ref{fig:fullSoluce}. This figure shows that the pre-accretion profile $X(P,t-t_0=0)$ (shown in black) is not step-like: the quiescent time from the previous cycle is not sufficient to reach a very small abundance of parent nuclei in the whole layer. After the accretion is switched on, the parent abundance profile increases (in dark blue) and reaches the stationary profile (in orange). At $t-t_0=3.8$\;years (dashed brown), the offset has already started and we have ${\dot{M}_a(t-t_0=3.8) = \dot{M}_{\rm max} /2}$: we observe that a decrease by half of the accretion rate barely changes the profile of $X(P)$. However, at $t-t_0=3.9$\;years (in teal), for which the accretion rate can be considered negligible with ${\dot{M}_a(t-t_0=3.9)/ \dot{M}_{\rm max} \simeq 10^{-5}}$, $X(P)$ has significantly evolved from the stationary profile.

The time needed to reach the stationary profile for a given reaction layer can be estimated from the formula in Eq.~\eqref{eq:taua} with $P_{\rm th}$ replaced by $\delta P$ defined by Eq.~\eqref{deltaPreac}:
\begin{align}
    \tau_{\rm stat} &=\tau_{\rm acc}\frac{\delta P}{P_{\rm th}} \;.
\end{align}
The formula is most accurate close to the threshold; therefore, we can defined the time needed to reach the stationary profile at $X=0.95$ and use the following formula:
\begin{align}
    \tau_{\rm stat}(X=0.95) &= 2.5 \bigg(\frac{1}{\bar{W}_1^5}{\tau_{\rm ec}\tau^3_{\rm acc}}\bigg)^{1/4}\nonumber \\
    &=0.4\,{\rm yr}\,R^3_{10}\tau^{1/4}_{\rm ec,5}W^{-5/4}_{10}\left(\frac{P_{\rm th,-5}}{\dot M_{-8} M/M_\odot}\right)^{3/4}
\end{align}
where $R_{10}=R/10{\rm km}$, $\tau_{\rm ec,5}=\tau_{\rm ec}/10^5{\rm s}$, $W_{10}=W/10{\rm MeV}$, $P_{\rm th,-5}=P_{\rm th}/10^{-5}{\rm Mev/fm^3}$, and $\dot M_{-8}=\dot M/10^{-8}M_\odot/{\rm yr}$. The timescale needed to reach the stationary state is on the order of the active phase (accretion) in low mass X-ray binaries. To decide whether this state is reached for a given reaction layer, one needs to study the full solution of the continuity equation.

To assess the role of the value of the maximum accretion rate on our results, we present in Fig.~\ref{fig:fullSoluceMdot} the parent nucleus abundance $X(P,t)$ for ${\dot{M}_{\rm max} = 10^{-10}\;{\rm M}_{\odot}}$ per year instead of ${10^{-8}\;{\rm M}_{\odot}}$ per year; the same accretion cycle parameters are used. The stationary profile for ${\dot{M}_{\rm max} = 10^{-10}\;{\rm M}_{\odot}}$ is different from the stationary profile for ${\dot{M}_{\rm max} = 10^{-8}\;{\rm M}_{\odot}}$ presented in Fig.~\ref{fig:stationary}. The solution $X(P,t)$ does not reach the stationary profile before the accretion is switched off. The profile of $X(P,t)$ evolves slower than for the value of the accretion rate $\dot{M}=10^{-8}$\;M$_{\odot}$ per year; the layer of electron captures is also much thinner. The accretion rate is directly related to the number of parent nuclei deposited onto the neutron star surface, and the speed at which nuclei are pushed to higher pressures. Therefore, this quantity dictates how fast the solution $X(P,t)$ increases towards the stationary profile.

The time of active accretion $t_a$ also plays a role on the thickness of the electron capture layer: from Fig.~\ref{fig:fullSoluce} with an accretion rate $\dot{M}=10^{-8}$\;M$_{\odot}$ per year, we can infer that if the accretion is switched off before or around 0.6\;years (blue curve in the figure), the stationary profile is not reached during the active phase period; in this case, the stationary state is reached after $\sim 1\,{\rm yr}$.

%-------------------------------------------------------------------------%
\subsubsection{Quiescence following a stationary profile}\label{sec:quies_sol}

We now consider the particular case of quiescence (${\dot{M}(t)=0}$) for the continuity equation Eq.~\eqref{eq:ce}:
\begin{equation}
    \frac{\rm d}{\rm dt }\ln \bigg( n\big(X(P,t),P\big)X(P,t) \bigg)  = - \mathscr{R}_{\rm ec} \big( X(P,t), P\big), \; \label{eq:continuity_quiescence}
\end{equation}
which follows a stationary profile, namely, with the initial condition ${X(P, t=0)=X_{\rm stat}(P)}$. Results for 40\;years of quiescence are presented in the movie available as Supplemental Material.

In the mixed layer approach, $X(P,t)$ in the quiescent phase is not zero in the shell, contrary to the instantaneous approach: electron captures still occur even when the accretion-related pressure increase is switched off. All $X(P)$ profiles evolve very rapidly during the first $\sim$ 10\;years after the beginning of quiescence. However, the profiles evolve increasingly slowly (similarly to an exponential decrease) in the remaining years of the quiescent phase towards a step-like function that is not actually reached.

%-------------------------------------------------------------------------%
\subsubsection{Accreted versus reactive baryons}\label{sec:reactivebaryons}

The number of accreted baryons $N_b$ (see Eq.~\eqref{eq:accBaryons}) and the number of baryons effectively involved in electron captures (reactive baryons) $N_r$ (see Eq.~\eqref{eq:reactiveBaryons}), are equal in the instantaneous approach, but not in the mixed layer approach. For the mixed layer approach, when considering only the stationary case of active accretion, we also have $N_b=N_r$: the stationary solution represents an equilibrium between the number of parent nuclei pushed beyond the threshold pressure of the reaction by continuous accretion and the number of parent nuclei transformed into daughter nuclei when they react deeper in the shell.

When considering the time-dependent solution $X(P,t)$ during the active phase, not all parent nuclei accreted necessarily have time to undergo their electron capture. For an active phase of four\;years with ${\dot{M}_{\rm max}= 10^{-8}}$M$_{\odot}$ per year, the number of accreted baryons is ${N_b = 4.3\times10^{49}}$. During the same period of active accretion, the time-dependent solution for $Z_0=26$ leads to $N_r \simeq 3.9 \times 10^{49}$: this corresponds to $\sim$ 90\% of the baryons accreted during the active phase having reacted.

During the quiescent phase, in the assumption that it follows from a stationary profile, the number of reactive baryons during quiescence is directly related to the speed of evolution of the $X(P)$ profile: we have ${N_r(Z_0=26) \simeq 6.1 \times 10^{48}}$, ${N_r(Z_0=24) \simeq 4.2\times 10^{48}}$, ${N_r(Z_0=22) \simeq 5.5\times 10^{48}}$ and ${N_r(Z_0=20) \simeq 2.1\times 10^{49}}$. Those number represent 14\%, 10\%, 13\%, and 48\% of the number of accreted baryons respectively for $Z_0=26, 24, 22,$ and $20$. For $Z_0=20$, the number of baryons that react during quiescence is not negligible compared to the number of reactive baryons during four\;years of the active phase in the stationary profile.

By adding the number of reactive baryons in the active phase and in quiescence for the full time-dependent solution for $Z_0=26$, we obtain $N_r^{\rm cycle} = N_b$ although a step-like function is not reached at the end of quiescence for the profile $X(P, t-t_q)$. Let us denote $N_{\rm nr}$ as the number of nuclei that have not reacted at the end of the quiescent phase. At the end of cycle ${i}$, we have $N_{\rm nr}^{i}$ parent nuclei that have not reacted in a small layer close to the reaction threshold (black line in Fig.~\ref{fig:fullSoluce}) of size $P_{\rm nr}^i-P_{\rm th}$. The value of $P_{\rm nr}^i$ and the number of nuclei that have not reacted during this cycle (proportional  to $P_{\rm nr}^i-P_{\rm th}$\footnote{The number of nuclei compressed above the pressure $P_{\rm th}$ that did not have enough time to react is $\sim \frac{4\pi R^4}{GMm_B}(P_{\rm nr}^i-P_{\rm th})$.} can be estimated comparing reaction the time ($1/ \mathscr{R}_{\rm ec}$) for $P_{\rm nr}^i$ with the quiescence timescale, $\tau_q$. For example, given the solution presented in Fig.~\ref{fig:fullSoluceMdot}, the value of ${\tau_q {\mathscr{R}_{\rm ec}}(P)}$ is on the order of 1 for $\bar{P} \sim 1.03$.

%-------------------------------------------------------------------------%
%-------------------------------------------------------------------------%
\subsection{Heat deposition in layers of electron capture}

%-------------------------------------------------------------------------%
\begin{figure}
    \centering
        \resizebox{\hsize}{!}{\includegraphics{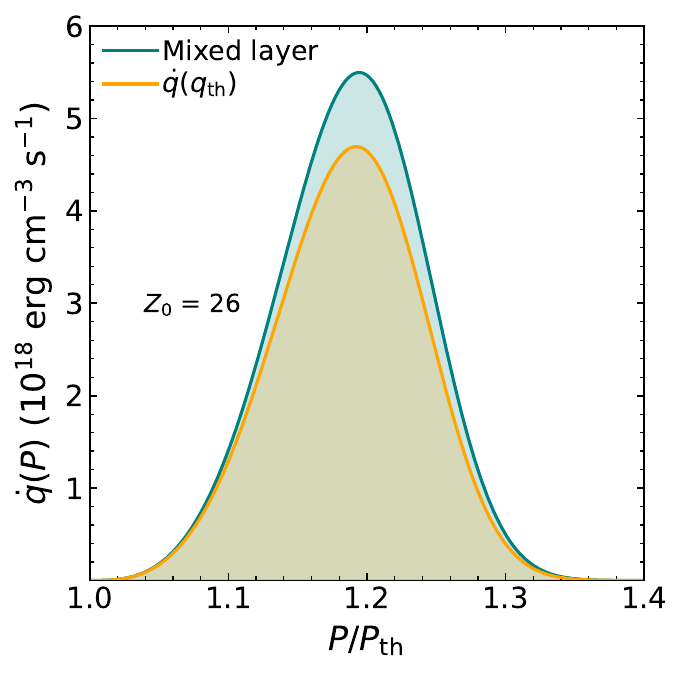}}
    \caption{Heating rate per unit volume $\dot{q}$ in the shell with proton number $Z_0=26$ as a function of the pressure during active accretion. This quantity is given for the mixed layer approach in blue and compared to the quasi-instantaneous approach in orange.}
    \label{fig:heatingrateCompare}
\end{figure}
%-------------------------------------------------------------------------%

\begin{table*}[]
    \caption{Heat release, neutrino emission and heat deposited in keV per accreted baryon for the four outer crust shells discussed in this paper, with parent proton number, $Z_0$. Results are presented for the stationary solution and for 40\;years of quiescence following a stationary profile.}
    \centering
    \begin{tabular}{|c|c||c|c|c|c|}
        \hline
        & Solution & $Z_0=26$ & $Z_0=24$ & $Z_0=22$ & $Z_0=20$ \\
        \hline \hline
        & & \multicolumn{4}{c|}{Heat release} \\ \cline{2-6} 
          \multirow{5}{*}{Stationary active phase} & Instantaneous & 38.9 & 41.8 & 34.0 & 24.3 \\ \cline{2-6}
        & \multirow{5}{*}{Mixed layer} & 47.5 & 42.5 & 34.5 & 25.1 \\ \cline{3-6} 
        & & \multicolumn{4}{c|}{Neutrino emission} \\ \cline{3-6} 
        &  & 36.0 & 32.7 & 26.2 & 19.0 \\ \cline{3-6} 
        & & \multicolumn{4}{c|}{Heat deposited} \\ \cline{3-6}
        &  & 11.5 & 9.8 & 8.3 & 6.1 \\ \cline{3-6} \hline \hline
        & & \multicolumn{4}{c|}{Heat release} \\ \cline{2-6} 
        \multirow{7}{*}{Quiescence following stationary} & Instantaneous & 0.0 & 0.0 & 0.0 & 0.0 \\ \cline{2-6}
          & \multirow{5}{*}{Mixed layer} & 6.5 & 4.2 & 4.4 & 12.1\\  \cline{3-6} 
         & & \multicolumn{4}{c|}{Neutrino emission} \\ \cline{3-6} 
         & & 4.9 & 3.2 &  3.3 & 9.1 \\  \cline{3-6} 
          & & \multicolumn{4}{c|}{Heat deposited} \\ \cline{3-6} 
         &  & 1.6 & 1.0 & 1.1 & 3.0 \\  \cline{3-6} \hline \hline
    \end{tabular}
    \label{tab:heatingNumbers}
\end{table*}
%-------------------------------------------------------------------------%

%-------------------------------------------------------------------------%
\subsubsection{Heat release in active phase: Stationary solution}\label{sec:stationary,active}

During active phases of accretion, the amount of heat released in the mixed layer approach and in the instantaneous approach are different. The reason lies in the pressure dependence, as demonstrated via Eq.~\eqref{eq:QabsMixed}: the chemical potential of electrons ensures that electron captures occurring at larger pressure (deeper in the layer) release more heat. In the instantaneous approach, all the heat is released exactly at the pressure threshold of the first electron capture and the heat release per parent nucleus is given by
\begin{equation}\label{eq:heatrateNoP}
    q_{\rm th} = W_1 - W_2 + E_{\rm exc} \;.
\end{equation}
In Fig.~\ref{fig:heatingrateCompare}, we present the heating rate per unit volume $\dot{q}(P)$ for the stationary solution of $Z_0=26$. To emphasize the role of the pressure dependence in the amount of heat released in layers of electron capture, we compare it to the heating rate per unit volume established from a "quasi-instantaneous" approach ${\dot{q}(q_{\rm th})}$ in which we assume that the energy release is equal to that at the threshold pressure. The pressure dependence of $q$ implies that the amount of heat released during the active phase is larger in the mixed layer approach than in the instantaneous approach. To illustrate this point, we computed the heat release per accreted baryon (see Eq.~\eqref{eq:totalQ}) for the stationary solution in the mixed layer approach for four years of active accretion with constant accretion rate. We then compared the outcome to the instantaneous approach, with the results presented in Table~\ref{tab:heatingNumbers} for $Z_0=26,\;24,\;22,$ and $20$.

The stationary profiles presented in Fig.~\ref{fig:stationary} represent the limiting case for the maximum potential heat release: $X(P)$ can never go beyond the stationary profiles at zero temperature. The stationary case corresponds to the maximum increase of the energy release due to the increase of the pressure and chemical potential at the point where most of the reactions take place compared to the instantaneous approach. This increase in $\delta E$ can be estimated from Eq.~\eqref{eq:QabsMixed} as ${\delta E = 2\delta\mu_e \simeq 0.5\mu_{\rm th} \delta P/P_{\rm th}}$ with $\mu_{\rm th}=W_1$. This approximation gives the increase in energy release (per nucleon) equal to $\sim 10,\,0.5,\,0.2,\,0.2$\;keV for $Z_0=26,\;24,\;22,$ and $20,$ respectively (on the same order as the numbers given in Table \ref{tab:heatingNumbers}).

%-------------------------------------------------------------------------%
\subsubsection{Heat release in quiescence following a stationary profile}

We compute the released heat during the forty years of quiescence phase that occur following a stationary profile and present results in Table~\ref{tab:heatingNumbers}. The heat release per accreted baryon by the deepest reaction layer ($Z_0 =20$) is quite large compared to other shells. When considering heat release per reactive baryons however, the values are very similar to what we see for the stationary heat release. For the deepest shell, the heat release per accreted baryon during quiescence is significant compared to the heat release during an active phase in the stationary profile. There are layers of electron captures present in the crust of accreting neutron stars for which the heat release during quiescence phases is quite large; hence, it is not negligible.

Finally, we note that the energy release in the instantaneous and mixed layer approach are on the same order of magnitude because the energy reservoir is directly connected to the mass (energy) difference of a nuclei under consideration. In this sense, the conclusions concerning the shallow heating problem where a much larger energy release is needed are the same as in the instantaneous approach (see e.g., Table 1 in \cite{Chamel2020}).

%-------------------------------------------------------------------------%
\subsubsection{Heat release in the active phase: Time-dependent solution for $Z_0=26$}

We computed the released heat per accreted baryon, by solving the full continuity equation during 4 years of active accretion followed by 40 years of quiescence with a time-dependent accretion rate following Eq.~\eqref{eq:mdotForm} and $\dot{M}_{\rm max}= 10^{-8}$\;M$_{\odot}$ in the case of $Z_0 = 26$. During the active phase,  we obtain a heat release of 42.9\;keV per accreted baryon. We have shown that the pressure dependence of Eq.~\eqref{eq:QabsMixed} leads to an increase in heat release $\sim 22$\% with respect to the instantaneous approach when considering stationary accretion in Section~\ref{sec:stationary,active}. However, over the same period of time in active phase, including a realistic time-dependent accretion cycle only leads to an increase of the heat release per accreted baryon by $\sim$~10\% when compared to the instantaneous approach. This is due to the decreased number of baryons that effectively react ($N_r=3.9\times 10^{49}$) compared to the number of accreted baryon ($N_b = 4.3\times 10^{49}$). 
If we compute the heat release per reactive baryons (and not per accreted baryons) for the time-dependent solution, we obtain 47.3\;keV per baryon during the active phase. This value is closer to the heat release per accreted baryon in the stationary case.

During quiescence, an additional $4\times10^{48}$ baryons are involved in electron captures, leading to a total heat release (active phase and quiescence) of 46.4\;keV per accreted baryon. This represents an increase of $\sim 19$\% compared to the heat release in the instantaneous approach.

%-------------------------------------------------------------------------%
\begin{figure}
    \centering
        \resizebox{\hsize}{!}{\includegraphics{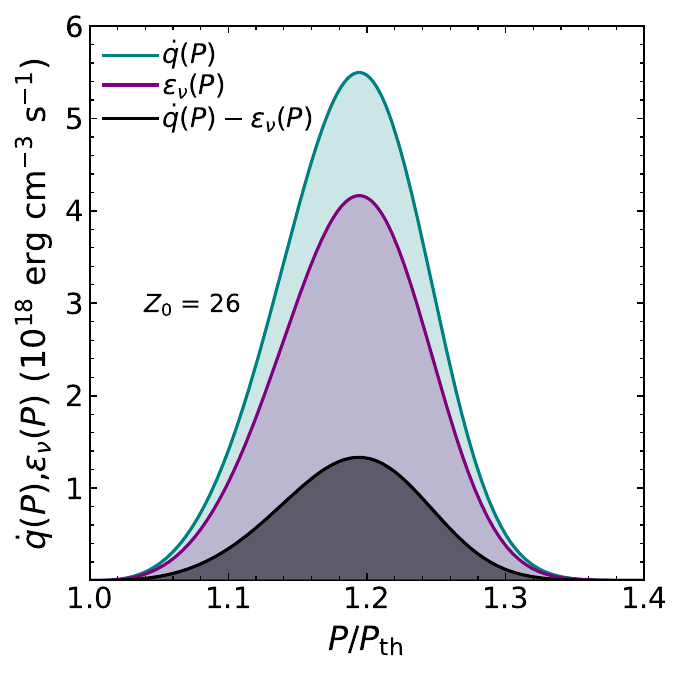}}
    \caption{Heating rate per unit volume $\dot{q}$ and neutrino loss 
    $\varepsilon_{\nu}$ as a function of the pressure established from the stationary solution of $Z_0=26$.
    }
    \label{fig:neutrinoCompare2}
\end{figure}
%-------------------------------------------------------------------------%

%-------------------------------------------------------------------------%
\subsubsection{Heat deposited}

Finally, in Fig.~\ref{fig:neutrinoCompare2}, we present the heating rate per unit volume $\dot{q}(P)$ and the neutrino loss rate $\varepsilon_{\nu}(P)$ for the stationary solution of $Z_0=26$. The neutrino loss (see Eq.~\eqref{eq:enuemission}) and total heat deposited (see Eq.~\eqref{eq:totalH}) for 4 years of stationary active accretion and for 40 years of quiescence with ${\dot{M}_{\rm max}= 10^{-8}}$\;M$_{\odot}$, are presented in Table~\ref{tab:heatingNumbers}. Overall, considerations of neutrino loss in an accreted outer crust with $^{56}$Fe ashes implies that the heat deposited is approximately one quarter of the heat released. The role of excited states of daughter and grand-daughter nuclei was discussed in \cite{Gupta2007}, who reported that the portion of heat release allocated to excitation energy compensated significantly the neutrino heat loss (see discussion in Section~\ref{sec:excited}). In the current paper, we only consider excited states for the daughter nucleus of the first electron capture.

%-------------------------------------------------------------------------%
%----------------------------- SECTION 4 ---------------------------------%
%-------------------------------------------------------------------------%
\section{Conclusion}

We studied the composition and heat release associated to electron captures in a fully accreted outer crust with $^{56}$Fe ashes when the kinetics of those reactions are taken into account. Making use of the mixed layer approach, we established the relation between the baryon density and the pressure of a lattice of nuclei permeated by a gas of degenerate electrons for a mixture of two nuclear species. Using recent nuclear physics experiments of nuclear mass measurements as well as the theoretical approach of Skyrme force based on the energy density functional described in \cite{Goriely2010}, we calculated the threshold  density and pressure for each of the four pairs of electron captures in our accreted outer crust. Nuclear data sheets, containing experimental data on nuclei lifetime with respect to $\beta$-decays, were used to derive the reaction rates of the four electron captures that dictate the kinetics relevant to our study. We explored the literature to find values for the accretion rate and typical durations for the active and quiescence phases that stem from observations of X-ray transient sources. We derived the continuity equation followed by the parent nuclei abundance and solved it numerically for realistic accretion parameters. The impact of electron capture kinetics on the composition in the shell was evaluated and we computed the thickness of the layers comprising both parent and grand-daughter nuclei for each pair of electron captures in the outer crust. Finally, we studied the impact of kinetics on the heat release, and  on the neutrino  heat loss in the shells. 

We find that the thickness in pressure for the layer containing a mixture of parent and grand-daughter nuclei of electron captures are on the order of $10^{-7}$\;MeV/fm$^3$ for the four pairs of electron captures in the outer crust. Our calculations were done neglecting thermal effects. Finite temperature would increase the reaction rates for the electrons due to the expansion of the available momentum space. Captures then become  possible even below the zero temperature threshold of the reaction
 \citep{Ushomirsky2000}, shifting the reaction layers to lower densities.

During the active phase of accretion, in the case of a constant accretion rate and studying the stationary solution, we find that including a finite reaction rate leads to an increase in heat release compared to the instantaneous approach that considers an infinite reaction rate for electron captures because electron captures can occur deeper in the shell. For a time-dependent active accretion of four years with a maximum accretion rate of $10^{-8}$\;M$_{\odot}$ per year, we find that the number of accreted baryons $N_b$ is larger than the number of baryons that are involved in electron captures $N_r$ during the active phase of accretion; this is contrary to the stationary solution, for which $N_b = N_r$. This effect can then significantly decrease the heat release per accreted baryon, as the time-dependent solution $X(P,t)$ does not remain in the stationary profile for the entire phase of active accretion.

We find that heat release occurs also during quiescence phases in the mixed layer approach (contrary to the instantaneous approach) and that said heat for some shells is not negligible, with respect to the heat released during the active phase. The neutrino loss significantly reduces the heat deposited in the matter, however, a more complete consideration of excited states is likely to compensate this loss. Those new results imply that the modeling of accreting neutron star thermal relaxation should be reevaluated by including time-dependent sources that would also cover the quiescent phases of the accretion process.

%-------------------------------------------------------------------------%
%----------------------------- SECTION 5 ---------------------------------%
%-------------------------------------------------------------------------%
\section{Acknowledgments}

We thank M. Fortin for useful discussions on accreting neutron star sources. We thank the referee for very pertinent remarks which have significantly improved the quality of the paper. The authors acknowledge the financial support of the National Science Center, Poland grants No. 2018/29/B/ST9/02013 and No. 2019/33/B/ST9/00942, as well as the National Science Foundation grant Number PHY 21-16686.

\bibliographystyle{aa}
% \bibliography{biblio}

%\newpage

\appendix 
\section{Derivation of the $\beta$-decay and electron capture rates}\label{AppendixA}

From Fermi's golden rule \citep{Dirac1927}, the probability of transition from an initial state denoted $i$ to a final state denoted $f$ is given by 
\begin{equation}
    \Gamma_{i\to f} = \frac{2\pi}{\hbar} |\mathscr{M}_{i \to f}|^2  \rho(E_f)\;,
\end{equation}
with $\mathscr{M}$ the matrix element of the reaction and $\rho(E_f)$ the density of final states with energy $E_f$. 

The matrix element for an electron capture $\mathscr{M}_{\rm ec}$ and its corresponding $\beta$-decay $\mathscr{M}_{\beta}$ are established from initial and final state products with the Hamiltonian of weak interaction denoted $\hat{H}_w$ such that 
\begin{align}
    &\mathscr{M}_{\rm ec} = \bra{\vec{p}_{\nu_e}, s_{\nu}}\bra{\vec{p}_{\rm f}, M_{\rm f}J_{\rm f}}  \hat{H}_w \ket{\vec{p}_{\rm i},  M_{\rm i}J_{\rm i}} \ket{\vec{p}_e, s_e}, \; \label{eq:ecnotation}\\
    &\mathscr{M}_{\beta} = \bra{\vec{p}_{\bar{\nu}_e}, s_{\bar{\nu}_e}} \bra{\vec{p}_e, s_e}\bra{\vec{p}_{\rm f}, M_{\rm f}J_{\rm f}} \hat{H}_w \ket{\vec{p}_{\rm i}, M_{\rm i}J_{\rm i}}, \label{eq:betanotation}\; 
\end{align}
with $\vec{p}_{\nu_e}$, $\vec{p}_{\bar{\nu}_e}$ and $\vec{p}_e$ the momenta of the neutrino, anti-neutrino and electron, $s$ designating the spin of leptons, and $M_fJ_f$ and $M_i J_i$ the nuclear spin and its projection on the spin quantization axis of final and initial nuclei. 

By averaging over initial states and summing over final states, the reaction rates denoted $\mathscr{R}$ are

\footnotesize{
\begin{align}
    \mathscr{R}_{\rm ec} &= \frac{2\pi}{\hbar} \Biggl \langle \sum_{\vec{p}_{\nu_e},\vec{p}_{Z-1}} \sum_{J_{Z-1}} |\mathscr{M}_{\rm ec}|^2 \delta(E_e - W - E_{\nu_e})  \Biggl \rangle_{\vec{p}_e, s_e, \vec{p}_Z, J_{Z}}, \; \\
    \mathscr{R}_{\beta} &= \frac{2\pi}{\hbar} \Biggl \langle \sum_{\vec{p}_{\bar{\nu}_e},\vec{p}_e,\vec{p}_{Z}}  \sum_{s_e}\sum_{J_{Z}} |\mathscr{M}_{\beta}|^2 \delta(W -E_e - E_{\bar{\nu}_e})  \Biggl \rangle_{ \vec{p}_{Z-1}, J_{Z-1}}, \;
\end{align}}
where the Dirac delta functions express the conservation of energy, and $E_e$, $E_{\nu}$, and $E_{\bar{\nu}}$ are the energy of the electron, of the neutrino and the anti-neutrino; for details on the absence of a sum over the (anti-)neutrino spin, we refer to \cite{Goldhaber1958}. 

In the heavy nucleus approximation, the transfer of momentum between the initial and final nuclei is neglected so that
\begin{align}
        \mathscr{R}_{\rm ec} &= \frac{2\pi}{\hbar} \Biggl \langle \sum_{\vec{p}_{\nu_e}} \sum_{J_{Z-1}} |\mathscr{M}_{\rm ec}|^2 \delta(E_e - W - E_{\nu_e})  \Biggl \rangle_{\vec{p}_e, s_e, J_{Z}} \;, \\
        \mathscr{R}_{\beta} &= \frac{2\pi}{\hbar} \Biggl \langle \sum_{\vec{p}_{\bar{\nu}_e}}  \sum_{\vec{p}_e} \sum_{s_e}  \sum_{J_{Z}} |\mathscr{M}_{\beta}|^2 \delta(W -E_e - E_{\bar{\nu}_e})  \Biggl \rangle_{J_{Z-1}} \;.
\end{align}
under the assumption that only allowed reactions are studied and that the lepton states are given by plane wave functions, the matrix element of each reaction can be simplified by operating a Taylor expansion of the lepton plane waves and keeping only the zeroth order term \citep{Povh2004}. The matrix element therefore does not depend on the momenta of leptons. Moreover, we make the assumption that the matrix elements relevant for the reactions do not depend neither on the spin of leptons nor on  the spin projection $M$ of nuclei. Consequently, we obtain such simplified  expressions
\begin{align}
        \mathscr{R}_{\rm ec} &= \frac{2\pi}{\hbar}  |\mathscr{M}_{\rm ec}|^2\Biggl \langle \sum_{\vec{p}_{\nu_e}} \delta(E_e - W - E_{\nu_e}) \sum_{J_{Z-1}}  \Biggl \rangle_{\vec{p}_e, s_e, J_{Z}}, \; \\
        \mathscr{R}_{\beta} &= \frac{2\pi}{\hbar} |\mathscr{M}_{\beta}|^2\Biggl \langle  \sum_{\vec{p}_{\bar{\nu}_e}} \sum_{\vec{p}_e} \delta(W -E_e - E_{\bar{\nu}_e})  \sum_{s_e}  \sum_{J_{Z}} \Biggl \rangle_{J_{Z-1}} \;.
\end{align}
Both the the  electrons and (anti-)neutrinos have a continuous spectrum, which translates the sum over their momenta into three dimensional phase-space integrals $I_{\beta}$ and $I_{\rm ec}$ that appear in the reaction rates and are given by 
\begin{align}
    I_{\rm ec} &= \frac{2\pi}{\hbar}  \int \frac{{\rm d}^3p_e}{(2 \pi \hbar)^3} \int \frac{{\rm d}^3p_{\nu}}{(2 \pi \hbar)^3} \delta\big(E_e - W - E_{\nu} \big) \;, \\
    I_{\beta} &= \frac{2\pi}{\hbar}  \int \frac{{\rm d}^3p_e}{(2 \pi \hbar)^3} \int \frac{{\rm d}^3p_{\nu}}{(2 \pi \hbar)^3} \delta\big(W - E_e - E_{\bar{\nu}} \big) \;. \label{eq:experimentalbetaf}
\end{align}
These integrals lead to the functions defined in Eq.~\eqref{eq:Fbfunction}.

\section{Calculating electron capture timescales}\label{sec:appendixTec}

In this section of the Appendix, we present details on how the values of the electron capture timescale presented in Table~\ref{tab:timescales} are established for each of the four reactions studied. Unless explicitly stated otherwise, experimental data from \cite{ENSDF} is used. 

%-------------------------------------------------------------------------%
\subsection{Reaction $^{56}$Fe $\to$ $^{56}$Mn}

In the $\beta$-reaction $^{56}_{25}$Mn $\to$ $^{56}_{26}$Fe, the allowed transition is  energetically prohibited, so that the inverse of this reaction produces an excited state $^{56}{\rm Mn}^*$ with a 110\;keV excitation energy. A doubly forbidden reaction is energetically favorable. However, the timescale of a $n$-forbidden reaction is much larger than that of an allowed reaction. Very roughly, the quantity $\log_{10}(ft_{1/2})$ is increased by some $2n+1$ compared to the allowed decay, see Sec.~ 2.1.4 of \cite{Grotz1990}. The timescale for $\gamma$-ray deexcitation of $^{56}$Mn$^*$ is less than  $10^{-9}$\;s, see Sec.~ 3.4 in \cite{Povh2004}. Moreover, we are concerned with the crust of an accreting neutron star which increases continuously the pressure in a piece of matter subject to electron capture. Assuming an accretion rate during the active phase $10^{-8}$\;M$_{\odot}$ per year, the time required for the chemical potential to reach the threshold of the allowed reaction is $\sim 0.3$\;years, which is small compared to the timescale of the doubly forbidden reaction. Therefore, we should consider an allowed reaction with additional heat (excitation energy) of $110$\;keV from the $\gamma$-decay.

Experimental data is available for the quantity $ft_{1/2}$ of the $\beta$-decay $^{56}_{25}$Mn$(3+) \to ^{56}_{26}$Fe$(2+)$, and we are interested in establishing the electron capture rate of the reaction $^{56}_{26}$Fe$(0+) \to ^{56}_{25}$Mn$^*(1+)$. We can now calculate the reaction timescale as
\begin{equation}\label{eq:reactionRateZ26}
    \frac{1}{\tau_{\rm ec}} = \frac{\ln(2)}{f t_{1/2}} \frac{21}{10} = 1.154\times 10^{-7} \; {\rm s}^{-1}\;, 
\end{equation}
with $f t_{1/2}= 10^{7.101}\; {\rm s}$\;.

%-------------------------------------------------------------------------%
\subsubsection{From $^{56}$Cr $\to$ $^{56}$V}
Experimental data is available for the quantity $ft_{1/2}$ of the $\beta$-decay $^{56}_{23}$V$(1+) \to ^{56}_{24}$Cr$(0+)$. We can define the inverse of the timescale of the electron capture as 
\begin{equation}\label{eq:reactionRateZ24}
    \frac{1}{\tau_{\rm ec}} = \frac{\ln(2)}{f t_{1/2}} \frac{9}{2} = 7.312\times 10^{-5} {\rm s}^{-1} \;,
\end{equation}
with $f t_{1/2} = 10^{4.63} \; {\rm s}$\;.

%-------------------------------------------------------------------------%
\subsubsection{From $^{56}$Ti $\to$ $^{56}$Sc}

Experimental data is only available for the quantity $T_{1/2}$ of the $\beta$-decay, the half-time for the decay of the nucleus through all channels of the reactions. Therefore, we assume that the channel we are studying, that is the $\beta$-decay $^{56}_{21}$Sc$(1+) \to ^{56}_{22}$Ti$(0+)$, is the dominant channel such that $T_{1/2} \simeq t_{1/2}$, with 
\begin{equation}
    T_{1/2} = 26 \;{\rm ms} \;. 
\end{equation}
The function $f$ which appears in the reaction rate of beta decays in Eq.~\eqref{eq:betaRate} is given by 
\begin{align}
     f &= \int_{1}^{\bar{W}} \sqrt{\bar{E}^2 -1} \bar{E} (\bar{E} - \bar{W})^2 {\rm d}\bar{E} \;. \label{eq:ffunction}
\end{align}

However, among the assumptions used on the matrix element of the $\beta$-decay, we have neglected the impact of the Coulomb interaction between electrons and proton on the shape of the wave functions of particles under consideration. To correct for that, it is customary to introduce the Fermi function (see, e.g., Sect.~2.1.3 in \cite{Grotz1990}).The non-relativistic formulation of the Fermi function is easier to implement than the relativistic one, however we have considered relativistic particles so far. To use the non-relativistic formulation, we have verified that the quantity $\log(ft_{1/2})$ calculated with the non-relativistic formulation of the Fermi function is close the the experimental data value for the electron captures involved in the first two shells of the accreted crust; we found a relative difference of at most 8\%, and therefore kept the non-relativistic formulation of the Fermi function in the evaluation of $ft_{1/2}$. For the $\beta$-decay $^{56}_{21}$Sc$(1+) \to ^{56}_{22}$Ti$(0+)$ reaction rate we obtain
\begin{equation}\label{eq:reactionRateZ22}
    \frac{1}{\tau_{\rm ec}} = \frac{\ln(2)}{f t_{1/2}} \frac{9}{2} = 1.362 \times 10^{-4} {\rm s}^{-1} \;,
\end{equation}
with $f t_{1/2} = 10^{4.36} \; {\rm s}$\;.

%-------------------------------------------------------------------------%
\subsubsection{From $^{56}$Ca $\to$ $^{56}$K}

Experimental data for this reaction is not available but calculations in the quasi-particle Random Phase Approximation are presented in \cite{Minato2021} for the quantity $T_{1/2}$ of the $\beta$-decay $^{56}_{19}$K $\to ^{56}_{20}$Ca. The energy level diagrams are not available either in \cite{ENSDF}, therefore we assume a $(0+) \to (0+)$ reaction. Similarly to the treatment of the previous reaction, we obtain the rate for the electron capture
\begin{equation}\label{eq:reactionRateZ20}
    \frac{1}{\tau_{\rm ec}} = \frac{\ln(2)}{f t_{1/2}} \frac{1}{2} = 6.307 \times 10^{-6} {\rm s}^{-1} \;.
\end{equation}
with $f t_{1/2} = 10^{4.74} \; {\rm s}$\;.

\end{document}